\documentclass[prl, twocolumn,showpacs]{revtex4} 
\usepackage{amsmath}
\usepackage{amsfonts}
\usepackage{amssymb}
\usepackage{graphics}
\usepackage{graphicx}

\def\d#1{#1^\dagger}

\begin{document}
\title{Experimental continuous-variable entanglement from a phase-difference-locked optical parametric oscillator}
\author{Jietai Jing}
\author{Sheng Feng}
\email[Now with the Center for Photonic Communication and Computing, Northwestern University.]{}
\author{Russell Bloomer}
\author{Olivier Pfister}
\email[Corresponding author: ]{opfister@virginia.edu}
\affiliation{Department of Physics, University of Virginia, 382 McCormick Road, Charlottesville, VA 22904-4714, USA}
\begin{abstract}
We observed continuous-variable entanglement between the bright beams emitted above threshold by an ultrastable optical parametric oscillator, classically phase-locked at a frequency difference of 161.8273240(5) MHz. The amplitude-difference squeezing is $-3$ dB and the phase-sum one is $-1.35$ dB. Besides proving entanglement in a new physical system, the phase-locked OPO, such unprecedented frequency-difference stability paves the way for transferring entanglement between different optical frequencies and densely implementing continuous-variable quantum information in the frequency domain. 
\end{abstract}
\maketitle 

The nondegenerate optical parametric oscillator (OPO) is a natural source of continuous-variable (CV)-entangled electromagnetic fields \cite{reid:1988}. Below threshold, it is a phase-sensitive amplifier whose quantum evolution can be described by a unitary two-mode squeeze operator \cite{walls:1995} which, in the ideal case, yields, for example, a common eigenstate of the amplitude-difference and phase-sum field quadratures. Amplitude and phase of a quantized field corresponding exactly to position and momentum of a mechanical quantum oscillator, this two-mode squeezed state is identical to that of the Einstein-Podolsky-Rosen paradox \cite{einstein:1935}, which has been implemented experimentally with finite squeezing \cite{ou:1992} and used in CV quantum information (CVQI) \cite{braunstein,braunstein:2005}. Above threshold, the OPO is a true oscillator rather than an amplifier and its dynamics become richer: as is well known, the phase difference of the two OPO signal beams undergoes, above threshold, an undamped diffusion process, driven by vacuum fluctuations and analogous to that of the phase of a laser beam, resulting in the Schawlow-Townes linewidth \cite{graham:1968}. There is, therefore, excess quantum noise on the phase difference of the OPO signal beams, compared to that of two independent ideal laser beams of the same power. This is a consequence of the number-phase Heisenberg uncertainty for the photon-number correlated OPO beams. We made the first experimental measurement of this excess quantum noise, which can also be understood as a macroscopic Hong-Ou-Mandel interference experiment \cite{feng:2004a}. It is, however, possible to suppress the Schawlow-Townes phase-difference drift by locking the phase difference of the signal beams of the OPO, thereby profoundly altering its natural dynamics and quantum properties. Indeed, perfect locking of the phase difference implies phase-difference squeezing, which means that the expected photon-number correlations in such a two-photon emitter are lost. This is clearly a different physical system from the standard OPO. Recently, CV entanglement was observed above threshold in standard OPO's \cite{villar:2005,su:2006} with unbridled Schawlow-Townes phase-difference drift. An elegant self-phase-locked type-II OPO, using polarization coupling from an intracavity waveplate, was demonstrated \cite{mason:1998} and theoretical studies predicted quantum properties very different from a regular OPO, as well as potential for entanglement generation for small values of the polarization coupling parameter \cite{adamyan:2004}. Experimental studies produced a record amount of CV entanglement below threshold \cite{laurat:2005a} but none above \cite{laurat:2005b}.

\begin{figure*}[htb]
\begin{center}
\begin{tabular}{c}
\includegraphics[width=1.4\columnwidth]{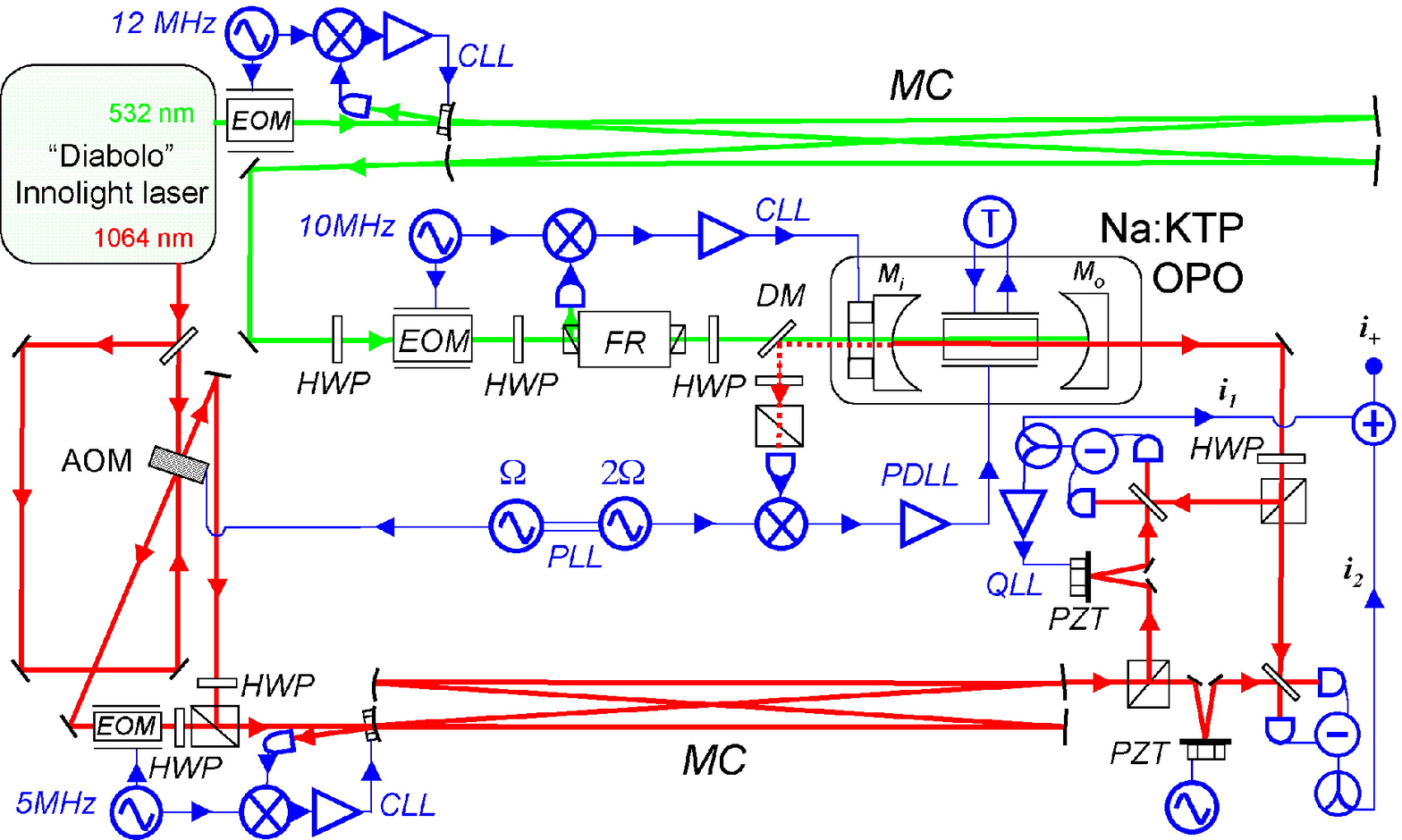}
\end{tabular}
\end{center}
\vspace{-.25in}
\caption{Experimental setup, see text. OPO mirror $\rm M_i$ has transmittivities T(532 nm) $= 0.98$ and T(1064 nm) $= 5\times 10^{-5}$; $\rm M_o$ has T(532 nm) $= 5\times 10^{-5}$ and T(1064 nm) $= 1.8\times 10^{-2}$. AOM: acousto-optic modulator. CLL: cavity-lock loop. DM: dichroic mirror. EOM: electro-optic modulator. FR: Faraday rotator. HWP: half-wave plate. MC: mode cleaner. P(D)LL: phase-(difference-)lock loop. PZT: piezoelectric transducer. QLL: quadrature-lock loop. The 12 MHz EOM is integrated in the laser.}
\label{setup}
\end{figure*} 
In this Letter, we report the observation of CV entanglement above threshold in a different type of phase-difference-locked OPO, in which the polarization coupling is derived from a classical beat note signal and applied electro-optically to the OPO nonlinear crystal. No theoretical model has yet been developed for this OPO. Unlike the aforementioned self-phase-locked OPO, the frequency difference of our OPO beams is not restricted to zero and can have any value within the phase-matching and electronics bandwidths. This is essential to enable entanglement and teleportation between different optical frequencies, e.g.\ an atomic resonance and the low-loss window of optical fiber. Phase-locking also suppresses uncontrolled phase and frequency drifts, which are detrimental to joint measurements of the quantum channel with external fields in teleportation. Finally, it opens the way to the fascinating regime of phase/frequency stable CVQI, combining the techniques of quantum optics with those of ultrastable frequency standards. One example is quantum heterodyne multiplexing, where multiple entangled mode pairs, with different frequency differences but the same frequency sum (as could be produced by a type-I OPO \cite{schori:2002}) can all be heterodyne-detected \cite{slusher:1985,feng:2004b} simultaneously using a single local oscillator, if their respective phase differences are locked \cite{jing}. Another example is the use of mode-locked optical oscillators and their frequency-comb spectrum as candidates for large-scale multipartite CV entanglement \cite{pfister:2004,bradley:2005}. 

\begin{figure}[htb]
\begin{center}
\begin{tabular}{c}
\includegraphics[width=.7\columnwidth]{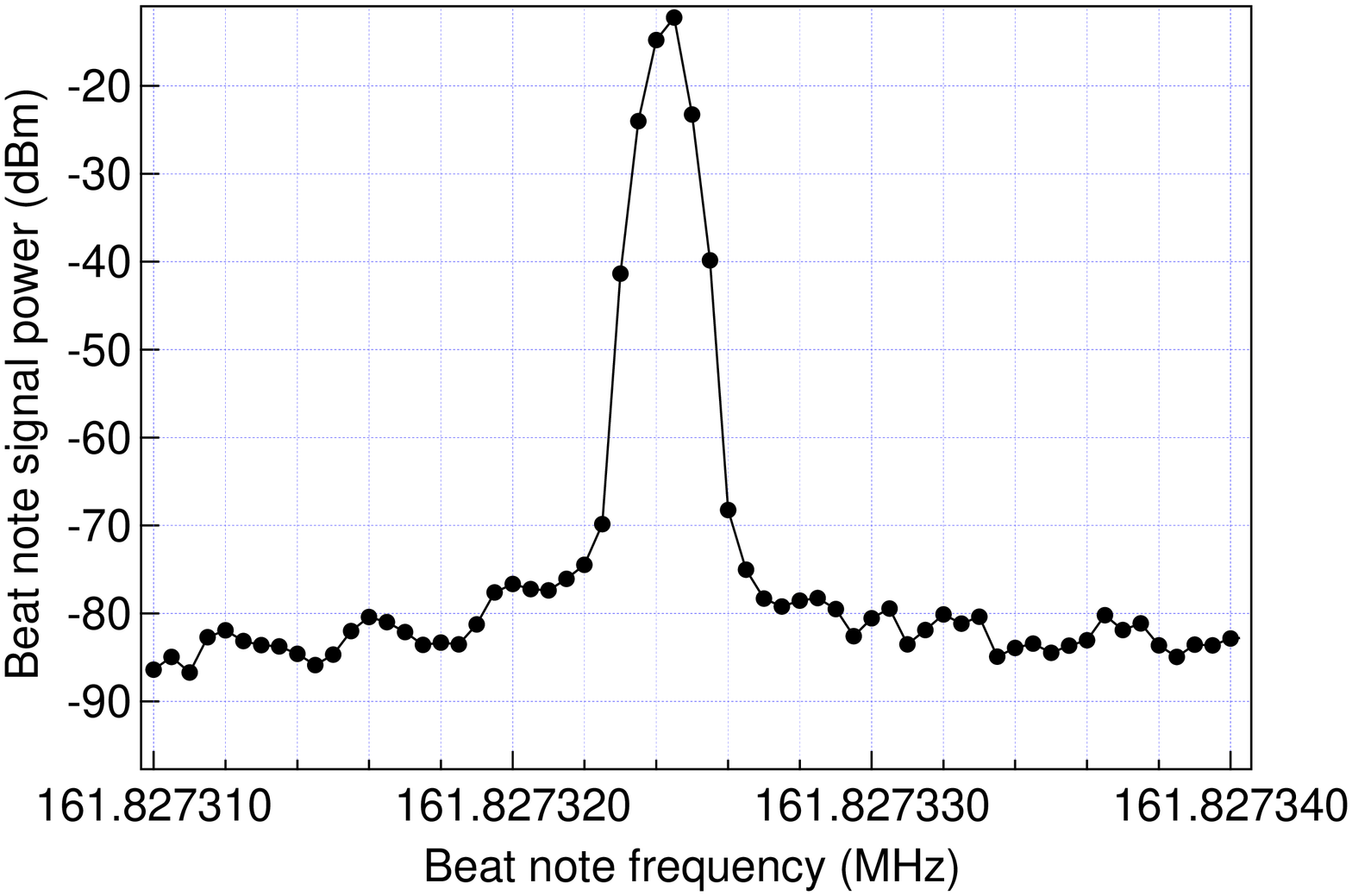}
\end{tabular}
\end{center}
\vspace{-.25in}
\caption{Ultranarrow  beat note of the phase-locked OPO. Resolution and video bandwidths are RBW = VBW = 1 Hz. Consecutive points are separated by 0.5 Hz. 100 averages.}
\label{beat}
\end{figure}

This experimental realization of CV entanglement of phase-locked bright CW beams used an ultrastable, doubly resonant, type-II near-concentric OPO based on an $X$-cut Na:KTP nonlinear crystal, temperature-stabilized at a few tenths of millidegrees, in which pump photons at 532 nm were downconverted into cross-polarized pairs at 1064 nm. This interaction was noncritically and collinearly phase-matched. The same OPO was used in our previous demonstration of macroscopic Hong-Ou-Mandel interference \cite{feng:2004a}. The experimental setup is sketched in Fig.\ref{setup}. The OPO pump and LO beams, at 532 and 1064 nm, respectively,  were provided by a Nd:YAG laser with an external resonant frequency-doubler (``Diabolo," Innolight). Both beams were spatially and temporally filtered by ``mode-cleaner" cavities, of respective half widths at half maximum (HWHM) 160 and 170 kHz. The twin OPO beams at 1064 nm exited through mirror $\rm M_o$, with typical operating powers from 1 to 10 mW (controlled by the pump power above its 65 mW threshold), and were separated by a polarizing beamsplitter (PBS). The reflected OPO-depleted pump beam was used as error signal of the OPO cavity lock loop (CLL). A weak leak at 1064 nm through $\rm M_i$ was picked off by a dichroic mirror and the resulting beat note of the twin beams was phase-locked to a stable synthesized radiofrequency at $2\Omega/(2\pi) = 161.827324$ MHz, by applying a correction voltage along the $Z$ axis of the Na:KTP crystal. With only the temperature lock and CLL, the frequency difference error was $\pm$150 kHz \cite{feng:2003}, due to large doping inhomogeneities in the crystal coupled to residual vibrations of the optical table. The phase-difference lock loop (PDLL) reduced this error by more than 5 orders of magnitude to less than 1 Hz (Fig.\ref{beat}), while keeping the frequency difference continuously tunable over tens of MHz. The individual OPO frequencies had a residual jitter of 10 kHz, measured by beating the OPO against the kHz-linewidth LO laser.
Note that the PDLL was purely classical, since its error signal was obtained from a balanced heterodyne measurement through a very low transmission mirror, both properties which independently cause the OPO quantum phase-difference noise to be replaced by vacuum fluctuations \cite{feng:2004a}. Hence, the PDLL phase-difference noise reduction could, in principle, degrade neither the conjugate amplitude-difference squeezing nor the entanglement. In order to observe the latter, i.e.\ the EPR correlation between the twin beams, we set up a standard double balanced homodyne detection (BHD) system, with low- and high-pass outputs. An acousto-optic modulator (AOM) was utilized to simultaneously up-shift and down-shift the frequencies of the LO, yielding two beams at frequencies $\omega\pm\Omega$ , where $\omega$  is the fundamental laser frequency and $\Omega /(2\pi)  = 80.913662$ MHz is the driver frequency of the AOM. The 1064 nm mode cleaner was built with a free spectral range of $\Omega/(3\pi)$ so as to allow simultaneous resonance of both frequency-shifted LO beams. The two cross-polarized outputs of this mode cleaner were suitable LOÕs for the two BHD systems since the frequency difference of the twin beams was phase-locked to $2\Omega$ by means of the PDLL. The two synthesizers working at $\Omega$  and $2\Omega$  were also electronically phase locked together, which suppressed any effect on the experiment of residual synthesizer frequency drifts. The optical part of the experiment was stabilized by 6 servo loops that controlled the OPO and mode-cleaner optical cavities (CLL), the OPO temperature (T), its phase difference (PDLL), and one of the LO phases, i.e.\ OPO quadratures (QLL), the other one being scanned for the purpose of data acquisition but lockable as well.

\begin{figure}[htb]
\begin{center}
\begin{tabular}{c}
\includegraphics[width=1\columnwidth]{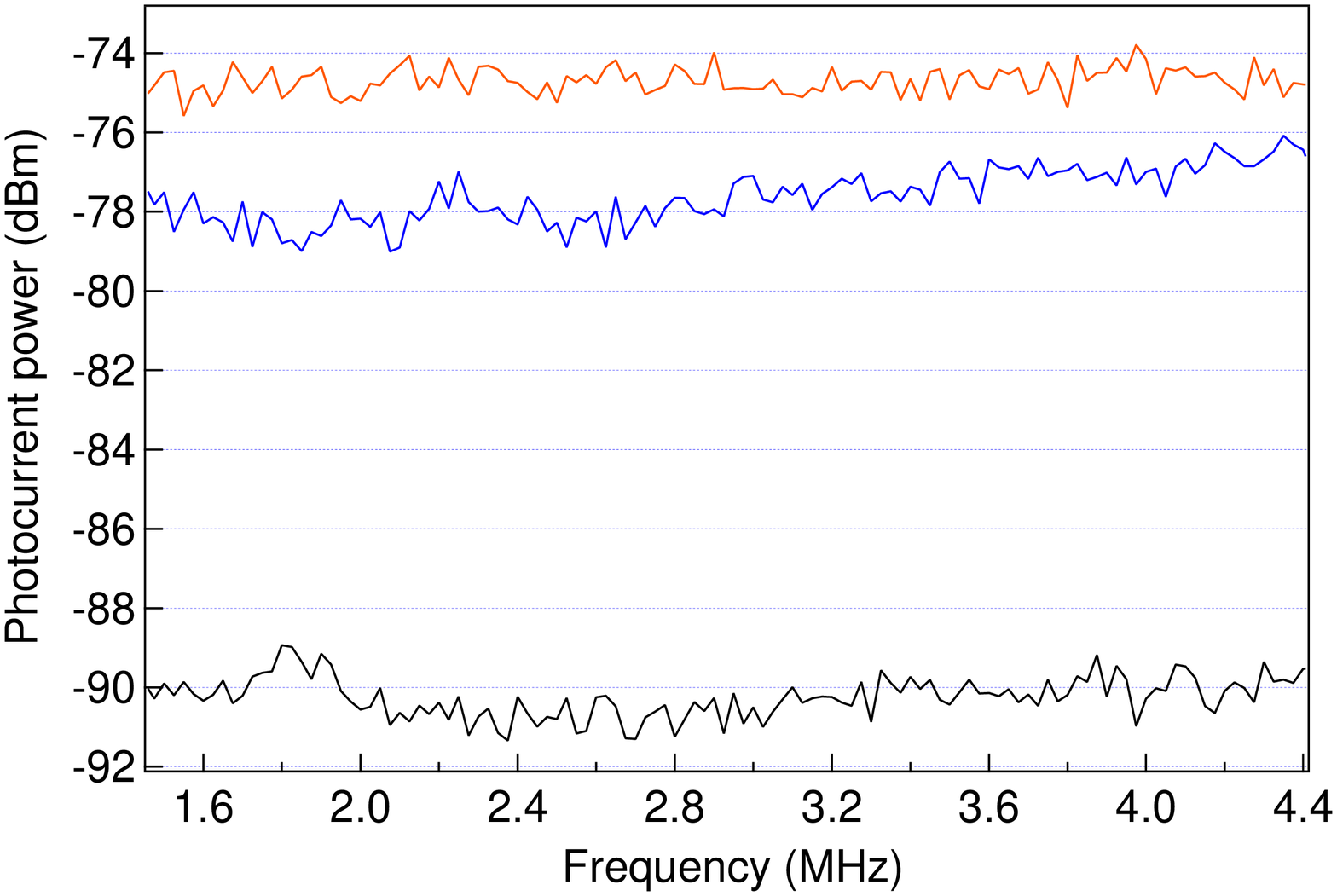}
\end{tabular}
\end{center}
\vspace{-.25in}
\caption{Intensity-difference squeezing. Red, shot noise level of both OPO beams. Blue, intensity-difference noise. Black, electronic detection noise. RBW=VBW=100 kHz, 100 averages.}
\vspace{-.1in}
\label{I-}
\end{figure}

\begin{figure}[htb]
\begin{center}
\begin{tabular}{c}
\includegraphics[width=1\columnwidth]{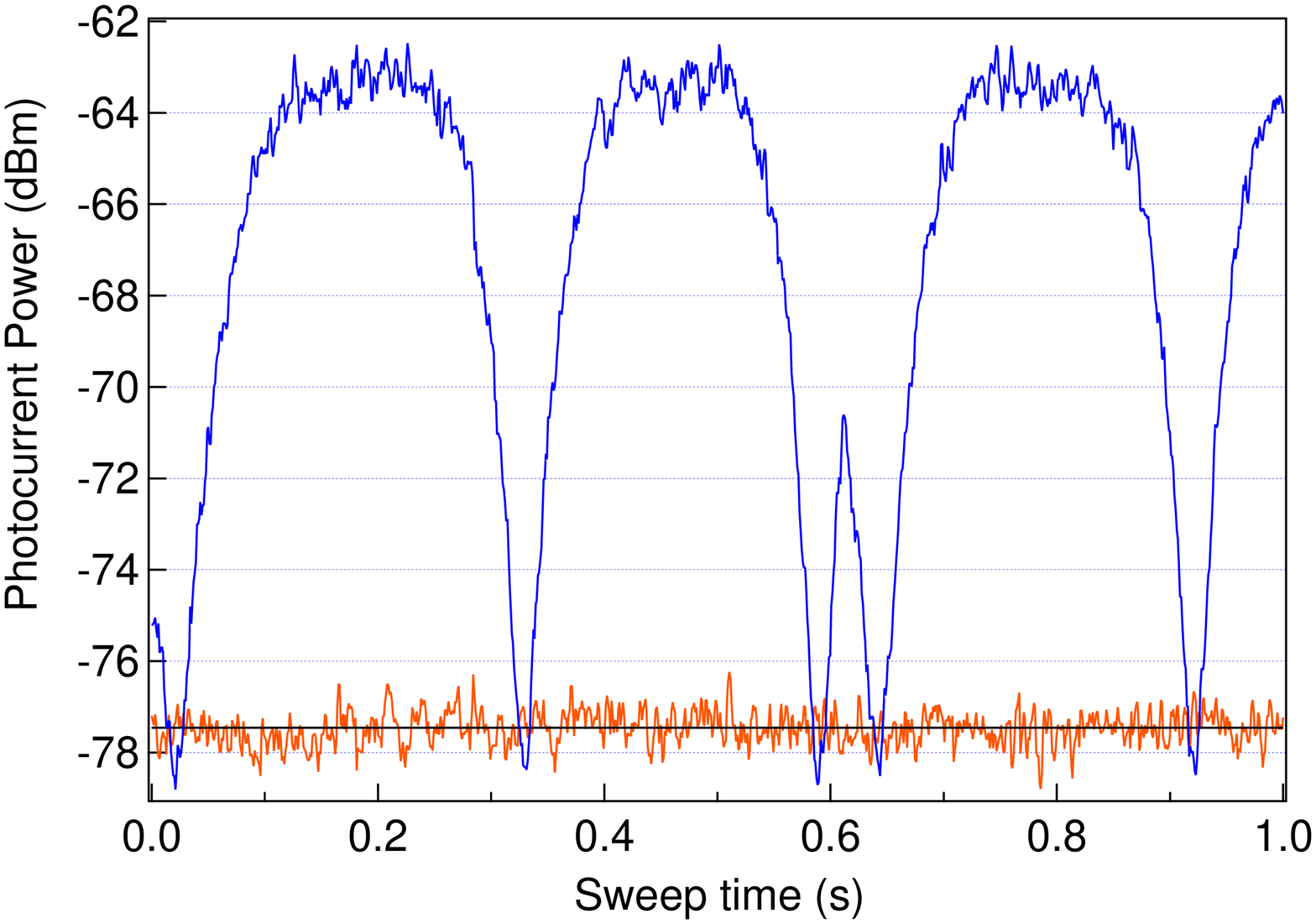}
\end{tabular}
\end{center}
\vspace{-.25in}
\caption{Phase sum squeezing (see text). Red, shot noise level of both OPO and LO beams. Blue, $\langle|\delta A_{1,\theta}(\Omega)+\delta A_{2,\pi/2}(\Omega)|^2\rangle$ versus $\theta$, at center frequency $\Omega/(2\pi)=1.7$ MHz. RBW 30 kHz, VBW 100 Hz, zero span, detection noise floor = -95 dBm. (Scan turning point at 0.6 s.)}
\vspace{-.1in}
\label{phi+}
\end{figure}

Figure \ref{I-} shows a typical intensity-difference squeezing spectrum of the OPO, measured by blocking the local oscillator beams, sending each OPO beam into a single photodiode, and electronically subtracting the photocurrents. The shot noise trace was obtained by rotating the OPO polarizations by $\rm 45^o$ before the PBS. Technical noise from the pump laser below 1.5 MHz prevented us from reaching squeezing levels stronger than $S_-=-3$ dB with respect to the shot noise limit (SNL) at 1.7 MHz. This amplitude noise can be reduced further by adopting a  mode-cleaner of HWHM closer to the laser linewidth, i.e.\ a few kHz.
Figure \ref{phi+} shows the quadrature sum noise of the twin beams versus one of the LO optical phases, the other LO phase being locked at $\pi/2$, i.e.\ to the phase quadrature. We verified on the DC interference fringe that the AC signal is squeezed only when the scanned quadrature is also the phase one, i.e.\ the phase shift is $\pi/2$.
The raw phase-sum squeezing is $-0.9$ dB (minimum squeezing hole from the shot noise average in Fig.\ref{phi+}). Moreover, one must take into account the fact that the bright OPO classical amplitudes beat with the (linearized) LO shot noise and mask the squeezing. Note that this is {\em not} squeezing degradation from optical losses and that self-homodyne detection  \cite{villar:2005,su:2006}, or an LO much more intense than the OPO beams, suppresses this problem. Let $a_{1,2}=\alpha +\delta a_{1,2}$ be the OPO photon annihilation operators and their fluctuations about a classical amplitude and $b_{1,2}=\beta+\delta b_{1,2}$ the same for the LO. We denote the Hermitian quadrature fluctuations of, say, $\delta a_j$ by $\delta A_{j,\theta}=e^{-i\theta} \delta a_j+e^{i\theta} \delta\d a_j$. The BHD photocurrent-sum fluctuations for a LO phase shift $\theta$ are given by the operator $\delta i_+ = \beta (\delta A_{1,\theta}+\delta A_{2,\theta}) + \alpha (\delta B_{1,\theta}+\delta B_{2,\theta})$, where the second term describes the aforementioned masking of the squeezing at $\theta=\pi/2$ by the LO shot noise. The raw measured squeezing in dB is $S_{\mathit{exp}} = 10 \log[\langle(\delta i_+)^2\rangle/\langle(\delta i_+)_{SN}^2\rangle]$ (decimal log). The OPO power was 2.8 mW per beam (pump power 5\% above threshold) and the LO power 6.5 mW per beam. We denote the ratio of these powers by $\rho$ and also take into account the nonideal quantum efficiencies of the photodetectors ($\eta=0.95$), and the BHD contrasts $C_1=0.986=\sqrt{\eta_1}$, $C_2=0.928=\sqrt{\eta_2}$. The true squeezing is thus
\begin{eqnarray}
S_+&=&10\log\left(\frac{2(\rho+1)[\rho(1-\eta)+1]}{\eta(\eta_1+\eta_2)}\ 10^{\frac{S_{+\mathit{exp}}}{10}}\right.\nonumber\\
&&\left.-\frac{2(\rho+1)+(\eta_1+\eta_2)[\rho(1-\eta)-\eta]}{\eta(\eta_1+\eta_2)}\right),
\end{eqnarray}
which yields $S_+=-1.6$ dB. Taking only into account the power factor $\rho$ ($\eta=\eta_1=\eta_2=1$) yields $-1.35$ dB. The theoretical value for the squeezing amount can be derived from \cite{reid:1988} in the ideal case and from a semiclassical analysis \cite{fabre:1989} including losses, and gives $-2.6$ dB at 1.7 MHz, given our cavity parameters. The cause of the 1 dB discrepancy may be uncorrected RF noise from the PDLL. This is thus a proof-of-principle demonstration. In order to improve it, one needs to better suppress the pump laser noise with a narrower mode cleaner and further optimize the PDLL filter, as PDLL and CLL are coupled \cite{feng:2004c}. This will allow one to work at lower signal frequency, well within the squeezing bandwidth. The reader will also have noticed that fringe contrast $C_2<C_1$, which substantially degrades the squeezing. This is due to an optical aberration in one of the OPO beams, stemming from the OPO crystal's natural anisotropy: one of the two OPO beam sections is elliptical whereas the other is circular. We observed that the ellipse's eccentricity increased with the length of the OPO's near-concentric cavity. We believe this is caused by walkoff in the wings of the focused beam in the Na:KTP crystal. Even though the propagation direction in the crystal is principal axis $X$ and should therefore give no walkoff, the beam is focused and its plane-wave angular spectrum does contain wave vectors at an angle with $X$. The $Z$ polarization experiences a strong birefringence in the extraodinary $XZ$ plane but not in the ordinary $XY$ plane. The $Y$ polarization, however, sees only weak birefringence in the extraordinary $XY$ plane, since Na:KTP, like KTP, is close to uniaxial ($n_X\simeq n_Y<n_Z$). Hence, only the $Z$ polarization acquires a significant mode-mismatch with the $\rm TEM_{00}$ LO mode. This could be corrected by inducing the same exact aberration on the LO mode, which is not trivial unless one uses the same OPO cavity to create the same eigenmode for the LO. To alleviate this problem, we defocused the beam inside the OPO cavity by reducing its length. We then measured the Gaussian beam parameters by measuring the intensity profile with a scanning pin-hole and found that the beam polarized along the $Z$ axis (horizontal) of the crystal had a horizontal-to-vertical waist aspect ratio of 1.29(5) whereas the $Y$-polarized (vertical) beam had a waist aspect ratio of 0.98(5). This had, however, the adverse effect of increasing the cavity eigenmode waist and therefore the OPO threshold from 15 mW \cite{feng:2003} to 65 mW, which lead to higher output power in the same operating conditions (number of times above threshold of the pump power), i.e.\ a larger power ratio $\rho$ as defined above and a consequent reduction of phase-sum squeezing. This issue can be alleviated by use of a stronger nonlinearity or of detectors that can withstand higher optical powers, both of which will yield a smaller $\rho$.

The aforementioned squeezing levels ($S_-$ = -3.0 dB, $S_+$ = -1.35\ dB) translate into  $\Delta(A_{1,0}-A_{2,0})= 10^{S_-/20} =0.71<1$ and $\Delta(A_{1,\pi/2} + A_{2,\pi/2})= 10^{S_+/20}=0.86<1$, which proves squeezed-state entanglement \cite{leuchs:2003}. Another quantitative measure of CV entanglement 
is the Duan-Simon criterion \cite{duan:2000,simon:2000}
\begin{equation}
\Delta\left(\frac{A_{1,0}-A_{2,0}}{\sqrt 2}\right)^2 + \Delta\left(\frac{A_{1,\pi/2}+A_{2,\pi/2}}{\sqrt 2}\right)^2 = 1.24 < 2.
\end{equation}
This is the strongest entanglement obtained to date for the beams emitted by an OPO above threshold.

In conclusion, we have demonstrated that the bright CW beams emitted by an electronically phase-locked nondegenerate OPO above threshold can be entangled and that phase-locking can, therefore, be used in a purely classical manner to yield the usual benefits of classical ultrastable operation for quantum information. There is no fundamental limitation to the frequency difference of the OPO beams and this result opens the way to quantum communication with intense ultrastable fields, as well as stable broadband two-mode and multimode \cite{pooser:2005} squeezing and entanglement for quantum communication and quantum heterodyne multiplexing. We thank Daruo Xie for discussions and Harvey Sugerman for the realization of electronic circuits. This work was supported by NSF grant Nos. PHY-0245032 and EIA-0323623.

\end{document}